\documentclass[preprint,aps]{revtex4}
\usepackage{graphics}
\usepackage{graphicx}
\usepackage{dcolumn}
\usepackage{bm}

\begin{document}

\title{Dynamic pinning at a Py/Co interface measured using inductive magnetometry}

\author{K.~J.\ Kennewell}
\author{M.\ Kostylev}
\affiliation{School of Physics, The University of Western Australia, 35 Stirling Highway, Crawley WA 6009, AUSTRALIA}

\author{M.\ Ali}
\affiliation{Department of Physics \& Astronomy, University of Leeds, Leeds, LS2 9JT UNITED KINGDOM}

\author{A.~A.\ Stashkevich}
\affiliation{LPMTM CNRS (UPR 9001), Universit$\acute{e}$ Paris 13, 93430 Villetaneuse, FRANCE}

\author{R.\ Magaraggia}
\affiliation{School of Physics, The University of Western Australia, 35 Stirling Highway, Crawley WA 6009, AUSTRALIA}

\author{D.\ Greig}
\author{B.~J.\ Hickey}
\affiliation{Department of Physics and Astronomy, University of Leeds, Leeds, LS2 9JT UNITED KINGDOM}

\author{R.~L.\ Stamps}
\affiliation{School of Physics, The University of Western Australia, 35 Stirling Highway, Crawley WA 6009, AUSTRALIA}

\begin{abstract}
Broadband FMR responses for metallic single-layer and bi-layer magnetic films with total thicknesses smaller than the microwave magnetic skin depth have been studied. Two different types of microwave transducers were used to excite and detect magnetization precession: a narrow coplanar waveguide and a wide microstrip line. Both transducers show efficient excitation of higher-order standing spin wave modes. The ratio of amplitudes of the first standing spin wave to the fundamental resonant mode is independent of frequency for single films.  In contrast, we find a strong variation of the amplitudes with frequency for bi-layers and the ratio is strongly dependent on the ordering of layers with respect to a stripline transducer.  Most importantly, cavity FMR measurements on the same samples show considerably weaker amplitudes for the standing spin waves. All experimental data are consistent with expected effects due to screening by eddy currents in films with thicknesses below the microwave magnetic skin depth. Finally, conditions for observing eddy current effects in different types of experiments are critically examined.
\end{abstract}

\maketitle

\textbf{1. Introduction}

Broadband ferromagnetic resonance microwave spectrometers have become a common experimental tool
with which to study dynamic properties of magnetic thin films and nano-structures \cite{Silva, Counil, Devolder1, Devolder2, Crew, self-organized, Schneider, Patton, Schneider2}. In this paper we demonstrate a unique ability of this technique for studying exchange effects in magnetic films and at buried interfaces in multilayer geometries. Resonance and standing spin waves are measured for Permalloy films and Permalloy/Cobalt bilayers, and we show how frequencies and amplitudes can be completely understood in terms of conductive layer microwave response.

Standing spin wave modes (SSWMs) are  excitations confined by the thickness of the film. The wavelengths of SSWMs are determined by the film thickness and pinning at the surfaces and interfaces. It is well known that the homogeneous microwave magnetic field typically used for ferromagnetic resonance (FMR) cavity experiments does not allow SSWM observation unless pinning
\cite{Kittel, Kittel1} or dynamic pinning \cite{Wigen} of magnetization is present at the film surfaces. Driving using a non-homogeneous field, e.g. by placing it on a hole in a wall of a microwave cavity \cite{Wolf}, can be used instead to observe the SSWM.
Recently it was shown theoretically that a microwave microstrip transducer can be used to couple efficiently to the SSWM\cite{kostylev}. In this scheme, resonant absorption by higher-order SSWM modes of any parity is predicted due to effects of eddy currents excited by the microwave field of stripline transducers. This in fact has allowed us to experimentally study the efficiency of coupling to these modes for an in-plane geometry using a broadband FMR and Network Analyzer technique (NA-FMR).

The plan of the paper is as follows. In the next section we discuss how a hmicrowave stripline transducer can drive standing spinwave modes in conducting multilayers. Results from experiments on single and bilayer structures are presented and discussed in the following two sections. The paper concludes with a discussion of circumstances under which conductivity can be expected to have significant effects in spin wave experiments.

\vspace{1cm}
\textbf{2. Excitation of precession of magnetisation  by microwave stripline transducers}

Results using two types of microwave transducers, coplanar and microstrip lines,  are discussed. The magnetic dynamics of thin magnetic films driven by the microwave field produced by the transducers can be understood by examining the quasi-static form of the Maxwell equations. We consider the geometry shown in Fig. 1 in which the static field and transducer axes are in the $z$ direction, and the normal to the film is in the $y$ direction. Written in terms of the Fourier-components of the microwave field, Maxwell's quasi-static equations are:
\begin{eqnarray}
 ikh_{ky}-\partial h_{kx}/\partial y& = \sigma e_{kz}	\\ \nonumber
 k e_{kz}& = -\omega \mu_0 (h_{ky}+m_{ky})	\\  \nonumber
 \partial h_{ky}/\partial y-ikh_{kx}& =-\partial m_{ky}/\partial y +ikm_{kx}.
\end{eqnarray}
 $\omega/(2\pi)$ is the driving frequency, $\sigma$ is the sample conductivity, and all components of the microwave field are presented as Fourier expansions in the in-plane direction  $x$ perpendicular to the transducer longitudinal axis $z$: $\textbf{m},\textbf{h},\textbf{e} \propto \int_{-\infty}^{\infty} \textbf{m}_k,\textbf{h}_k, \textbf{e}_k exp (i \omega t) exp(-ikx)dx$. These expansions are needed in order to describe the in-plane inhomogeneous microwave field of the stripline transducers. Equations (1) can be reduced to a single second-order differential equation \cite{Kim}:
\begin{equation}
\partial^2 h_{kx}/\partial y^2-[i \mu_0 \sigma \omega+k^2]h_{kx}=[i \mu_0 \sigma \omega-k^2]m_{kx}+ikm_{ky}.	
\end{equation}

An analytical solution to this equation in the form of a Green's function was obtained in \cite{Kim}. Here we briefly discuss essential peculiarities which follow from Eq.(2) for conductive films. 

Waves with non-zero $k$ in the film plane are traveling spin waves excited by the inhomogeneous field created by the transducers \cite{Counil}. The total microwave magnetic field in the samples  $\textbf{h}_{k}$ in Eqs. (1) and (2) consists of several contributions. The first contribution is the field of the stripline transducer $\textbf{h}_{ek}$. Another is the screening field $\textbf{h}_{Oek}$ which is also created by eddy currents in any \textit{nonmagnetic} conducting body on whose surface a microwave magnetic field  $\textbf{h}_{ek}$ is incident. The rest are magnetic contributions: the dipole $\textbf{h}_{dk}$ and the effective exchange $\textbf{h}_{exck}$ fields of the precessing magnetisation, and the field $\textbf{h}_{prk}$ which is created by the eddy currents which are induced by magnetisation precession.  The contributions $\textbf{h}_{ek}$ and $\textbf{h}_{Oek}$ to $\textbf{h}_{k}$ remain present in Eq.(2) even when $m_{kx}$ is set to zero, thus they represent the excitation field which drives magnetization precession.

In transducer experiments conductivity effects can be significant for  films thinner than a magnetic skin depth, $l_{sm}$ and can considerably modify efficiency of excitation of magnetisation precession by an external microwave field \cite{kostylev}. For the remainder of this paper, we refer to films satisfying this condition as 'SSDF' (an acronym for 'sub-skin depth films'). We present additional considerations related to this skin depth in the Conclusions, with particular reference to Almeida's approach for dipolar spinwaves \cite{Almeida}.

We note that strong microwave screening effects were predicted for thin films related to the phase of reflection of the microwave field from the film surface. The theory was constructed for a particular case of ferromagnetic resonance driven by very wide microstrip transducers ($k_{max}=0$). One may suppose that similar effects should have been noticed in the travelling wave experiments as well. In the following we show that it is actually not the case, and the broadband microstrip FMR represents a unique tool to observe these effects.

The microwave magnetic field outside the film is described by Eq.(2) with a vanishing right-hand part and $\sigma=0$.
This field is a combination of the microwave field from the transducer and the dynamic magnetic field from the film. Solving this equation outside the film, and applying the usual electromagnetic boundary conditions, one  arrives at  conditions for the fields at the film surfaces involving dynamic quantities \textit{inside} the film only.
The conditions at the film surface $y=L$ (not facing the microwave flux from the transducer) are:
\begin{equation}
k e_{zk}+i \omega \frac{\mid k \mid}{k} h_{xk}=0 .
\end{equation}

For the film surface facing the microwave flux $y=0$, the condition is:
\begin{equation}
k e_{zk}-i \omega \frac{\mid k \mid}{k} h_{xk}=i \omega h_k^{(0)},
\end{equation}

where $h_k^{(0)}$ is some quantity proportional to the stripline transducer field $\textbf{h}_{ek}$.

From (3) and (4) it follows that for $k=0$ the field $h_{xk}(y=0)=const\neq 0$, but $h_{xk}(y=L)=0$. This means that a very efficient shielding by microwave eddy currents takes place for SSDF films\cite{kostylev}. For $k\neq 0$ the shielding is not perfect and one can expect a significant microwave magnetic field at the rear film boundary.

The effects of conductivity have been previously considered also by Almeida and Mills \cite{Almeida} in the limit of exchange free, dipolar spin waves and $L>>l_{sm}$. Based on Eqs.(3-4) we extended this theory on smaller $L$-values. Results from a calculation using the extended theory are shown in Fig. 2. The value at the rear film boundary is shown  as $h_{xk}(y=L)/h_{xk}(y=0)$ vs. $k$ in Fig.(2b). From this figure one finds that the efficient shielding disappears at a $k$-value $k_s$ about 200 rad/cm (seen as $h_{xk}(y=L)/h_{xk}(y=0)=0.65$ for this $k$-value). One can relate this to disappereance of the contribution of the screening field $h_{Oek}$ to the total dynamic magnetic field of the film.

In summary, for the broad transducers considered in the present paper, the homogeneity of the microwave field in the film plane, together with film thicknesses small compared to the microwave skin depth, form conditions for observation of pronounced eddy current effects. We present below experimental evidence for eddy current effects in what follows, and show that these effects provide quantitative descriptions for observed intensities.

\vspace{1cm}
\textbf{3. Experiment}

 Measurements are made with the sample placed on a section of the microwave transducer carrying an AC microwave current as illustrated in Fig. 1.  The magnetization is aligned along the axis of the coplanar stripline by an applied static DC bias field.  Current through the transducer generates an oscillating magnetic field in the sample perpendicular to the equilibrium magnetization. Resonant absorption is detected by measuring transmitted intensities through the transducer.  Transmission measurements were made with very little attenuation from the cables or waveguide.  The strength of the bias field is swept from 0 to 0.6 Tesla, at a fixed frequency of the ac current. This is repeated across the available range of frequencies (100 MHz to 20 GHz).  In this way frequencies are chosen which avoid any waveguide resonances or sample reflections.  The magnetic contribution of the signal is extracted by measuring a reference signal at a field large enough (1 T) to suppress any of the resonances.  This range allows an optimal compromise between choosing thicker films in order to detect low frequency SSWM modes and measuring surface effects that fall off according to $1/t$.

The samples were deposited by magnetron sputtering at an argon working pressure of 2.5 mTorr.  Two series of samples were grown, each containing bilayer and single layer films. Details of each series are listed in Table 1. All samples in a given series were grown during the same vacuum cycle in order to ensure consistency. The base pressure prior to the deposition was of the order of $1 \times 10^{-8}$ Torr. The film structures Ta(5 nm)/Py($t_{Py}$ nm)/Co($t_{Co}$ nm)/Ta(2.5 nm (Series 1) or 5 nm (Series 2)) were deposited onto silicon (100) substrates in an in-plane forming field of magnitude 200 Oe at ambient temperature. Deposition rates were determined by measuring the thickness of calibration films by low angle x-ray reflectometry.

Note that the capping and the seed layer have the same thickness in Series 2, in contrast to the structures grown in Series 1. This gave Series 2 a symmetric combination with respect to transducer coupling to the film and substrate through the capping/seed layers. Comparison of the FMR data obtained on both series reveals no effect of change in the capping layer design.

All bilayers of Series 1 consist of a Cobalt (Co) layer grown on top of a thicker Permalloy (Py) layer ("Si/Py/Co" geometry.) Thicknesses of each Co and Py layer are varied across the series. All bilayers of Series 2 consist of a 10 nm thick Co layer and are divided into two sub-series.  The "Si/Py/Co" sub-series has the Cobalt layer grown on top of the Py layer as Si/Ta/Py/Co/Ta, similar to the structures in Series 1. The "Si/Co/Py" subseries has reversed ordering of Py and Co layers: Si/Ta/Co/Py/Ta. The total sample thickness for all samples in  Series 1 and Series 2 is smaller than the microwave magnetic skin depth \cite{Almeida} over the frequency range 100 MHz - 20 GHz. Note that the microwave magnetic skin depth varies from 102 nm at 7.5 GHz to 111 nm at 18 GHz as calculated from the material parameters derived from the single layer reference Py films, and assuming 0.008 for the Gilbert damping parameter and of $4.5 \cdot 10^6$ Siemens/m for the conductivity of Py \cite{permalloy}. Whereas the classical skin depth actually decreases with increasing frequency, the \textit{magnetic} skin depth instead increases slightly because of increased magnetic losses $\alpha \omega$ (see Eq.(2.9) in Ref.\cite{Almeida}).

An Agilent N5230A PNA-L microwave vector network analyzer (VNA) was used to apply the microwave signal to the samples and to measure magnetic absorption. As a measure of the absorption we use the microwave scattering parameter $S21$ \cite{Counil}. The sample sits on top of the transducer with the magnetic layers facing it. To avoid direct electric contact the sample surface is separated from the transducer by a 15 $\mu$m thick Teflon layer.  The microwave frequency is held constant and the static magnetic field $H$ is slowly increased. The raw data then appears as resonance curves in the form of $S21(H)$ as a function of $H$. This process is repeated for a number of frequencies. We also measure $S21$ for the transducer with no sample ($S21_0(H)$) to eliminate any field-dependent background signal from the results. The results presented below are $Re(S21(H)/S21_0(H))$. It is worth noting that the raw data $\mid S21(H) \mid $ show the same qualitative behavior, so artefacts arising from the mathematical processing of data are not significant.

Detailed broadband FMR measurements on the samples from Series 1 were taken using a coplanar transducer and compared to additional results made with a microstrip transducer.  One  measurement run at a single microwave frequency requires about 15 min of VNA time, a scan across the entire frequency range together with a background scan requires a day. These large time requirements limited the number of cases examined and only two frequencies were studied across the entire range using the microstrip transducer. The microstrip results were in complete agreement with those from the coplanar waveguide. The measurements on Series 2 were made using the microstrip transducer only, allowing also data to be taken for two sample orientations: one with the film facing the transducer and one with the substrate facing the transducer. Moreover, the microstrip transducer turned out to be of a slightly better microwave quality than the coplanar one.

Two additional measurements were made in order to make comparison with very different techniques. One set of data was collected using a section of a hollow waveguide in transmission and reflection, and a Varian 4 cavity in reflection. Another set of results was collected using a Brillouin light scattering technique.

\vspace{1cm}
\textbf{4. Single-layer films}

Results from the single layer reference Py films in Series 2 are shown in Fig. 3 for driving at 7.5 and 18 GHz.  A fundamental resonance and a standing spin wave mode can be identified for each film except the thinnest 30 and 40nm-thick films at 7.5 GHz. For these films this frequency is lower than the minimum frequency for observation of the first SSWM.

The strongest absorption peak is identified with the fundamental resonance mode. The resonant field for this mode is independent of Py thickness, suggesting that surface anisotropies can be neglected.
The smaller peak is identified with the first (odd symmetry) SSWM. Since there is no evidence for surface anisotropies, the SSWM is assumed to be unpinned. The resonant fields for this mode are well described using our theory \cite{kostylev} with an exchange constant of $A=6 \cdot 10^{-7}$ erg/cm, a gyromagnetic constant of 2.82 MHz/Oe, and a saturation magnetization for Py of $4 \pi M_s=$8320 G. Details of the fitting procedure are given at the end of this section.

The value for $4 \pi M_s$ derived from the broadband FMR measurements is consistent with results obtained with SQuID magnetometry from the 74nm thick reference sample of Series 1. A saturation magnetization for the Py given by $4 \pi M_s=$8000 G was obtained from out of plane saturation and volume magnetization measurements.
This value is in good agreement with the value shown above extracted from fitting the FMR data.

A comparison of the VNA FMR data with results from cavity FMR measurements taken at 9.47 GHz is shown in Fig. 4 (Panels a-c). One sees that the SSWM's are clearly visible in the broadband FMR but appear very weak in the cavity FMR experiment. We expect that the unpinned SSWM should not produce a strong response in the cavity due to the high symmetry of the mode, resulting in a low overlap integral with the highly homogeneous driving field. Nevertheless, some signature of this mode is apparent in all the films from this series. This suggests that a weak surface  pinning may be present and must be asymmetric as well; i.e., larger at one film surface than the other.

As a check to confirm that the smaller peak seen in the broadband FMR is in fact the lowest frequency odd symmetry SSWM, a Brillouin light scattering (BLS) study of the 60nm thick single-layer film was performed. In thermal BLS SSWM's are excited through thermal fluctuations and can therefore appear regardless of mode symmetry \cite{Grimsditch}. A BLS intensity spectrum measured at the angle of light incidence of 5 degree from the normal to the film is shown in Panel 3(d).  We found that the frequency position of the first BLS peak above the fundamental (F) dipole mode for a given field and a given magnon wavenumber is consistent with the field position of the lower field peak detected in the broadband FMR. This unambiguously identifies the smaller peak as the first odd symmetry SSWM (SSWM1 in the panel). Furthermore, the BLS data indicate the presence of a dipole gap \cite{stashkevich} in the spin wave spectrum, where the Damon-Eschbach mode repels the first dipole exchange  branch \cite{de wames}. Without pinning, the modes are orthogonal and should cross, and the existence of a gap indicates some small degree of asymmetric pinning at the film surfaces that modifies the symmetry of the modes. Full details of the BLS study will be published elsewhere.

Lastly, we comment on eddy current effects on the transducer FMR response. Amplitudes for absorption are well described by the theory of Kostylev \cite{kostylev}. In particular, it is shown that large amplitude SSWM peaks can be observed from metallic films without surface magnetisation pinning provided that an asymmetric eddy-current contribution to the total microwave magnetic field exists. This can in fact be realized simply by stripline transducers and used to detect SSWM resonances that are weak or not visible in conventional cavity FMR. This eddy current theory also describes, as experimentally observed, that the SSWM amplitudes in single-layer films are practically independent of driving frequency.

A detailed study of SSWM's was carried out using the reference 74nm-thick single-layer Py film from Series 1. A coplanar transducer was chosen to drive and detect magnetization precession. Measurements were taken with the static applied field aligned along the direction of the uniaxial anisotropy axis, and field sweeps were made from 0 up to 15GHz. The measured dependencies of resonance fields for the observed modes on the driving frequency are shown in Fig. 5a. For this film a best fit gives a saturation magnetization value of $4\pi M_s=8120\pm60$ G, g-factor of $g=2.05$, and an exchange constant of $A \simeq 0.41 \times 10^{-6}$ erg/cm.  These parameters provide consistent results for all measured Py thicknesses for Series 1.

The fits were made as follows. Since it was found that conductivity and an eventual weak surface pinning of magnetization have negligible effect on the frequency of the fundamental mode, the $\omega(H)$ dependence for this mode was used in the expression for in-plane FMR, $\omega^2=\gamma^2H(H+M_s)$, in order to determine $M_s$. A best fit was obtained using regression analysis. The values of $M_s$ obtained this way were then used to fit the experimental data for the SSWM frequency $\omega/(2 \pi)=15$GHz with our theory \cite{kostylev}, leading to the above value of A.

As was done for data from Series 2, we have assumed that the SSWM peak corresponds to the first standing spin wave mode. As noted earlier, efficient excitation of the asymmetric modes is not possible unless a film has very different pinning conditions for magnetization at two film surfaces. Using the approach of Ref. \cite{dmitriev-kalinikos} we find that the experimental intensity for the SSWM cannot be obtained unless one assumes a near complete pinning of magnetization at one of the film surfaces. Such a large pinning is inconsistent with the small cavity FMR mode intensities. Strong pinning is also inconsistent with measured resonant field since pinning considerably shifts the fundamental and SSWM resonant fields downwards. Moreover, if one assumes that the SSWM is instead the second, symmetric, SSWM one then obtains an unrealistically small value for the exchange constant: $A \simeq 0.11 \times 10^{-6}$ erg/cm. Furthermore, as will be discussed in the next section, the  spectra from bilayer films in this series display several high order SSWM's. We therefore conclude that the observed SSWM is in fact the first, odd symmetry mode, and is not strongly pinned at either surface. This observation is thus additional evidence for eddy current induced SSWM in conducting films.

\vspace{1cm}
\textbf{5. Bi-layers}

Resonance curves for bilayers from Series 2 are shown in Fig. 6. Except for panels 5a and 5e, each plot contains a response from a Si/Py/Co bilayer and a response from a Si/Co/Py structure. The Py layer thickness is the same for both orderings in each panel. One sees that the response of the Si/Co/Py structures is characterized by a single absorption peak located at a field slightly little smaller than that of the fundamental mode for the corresponding Py single-layer. We will denote this single-peak response as "Type A". The field downshift decreases as the Py layer thickness is increased, indicating a dynamic magnetization pinning. The reversed ordering of layers (Si/Py/Co) provides a quite different response, and we denote this as a "`Type B"' response. Two peaks with comparable intensities are seen at 7.5 GHz in panels 1b to 1d. The high field peak is located at the same field as the Type A response, and is identified as the fundamental. The low field peak is at a field close to the first SSWM observed in the single-layer film, and should therefore be the first SSWM mode for the bilayer. A field downshift is observed that decreases with increase in Py thickness, and thereby indicates dynamic pinning \cite{Wigen}.

The inset to Panel (d) shows cavity FMR data for the Si/Py/Co sample with 80 nm of Py, and 10 nm of Co. The measurements were taken at 9.47 GHz. Three distinct peaks are seen. The largest peak is the fundamental mode and the two other peaks are the 1st and the 2nd SSWM. From this figure inset two important facts can be determined. First, additional SSWMs are detected that are not visible in the microstrip broadband FMR data from the Si/Co/Py structure. Secondly, the  SSWM amplitudes are considerably smaller than corresponding ones found for the Si/Py/Co structure.

Coplanar-transducer studies were carried out for Series 1. As with the Si/Py/Co sub-series 2, only Type B responses were detected for the Si/Py/Co sub-series 1. The mode spectrum of the bilayers exhibit no observable effects with the addition of between 0 and 1 nm of Co. This may be understood as partial coverage by the Co such that unconnected or weakly connected clusters exist on the Py. The measurements are made at room temperature, and it may be that the clusters are too small to form a stable magnetic moment, or form appreciable magnetocrystalline anisotropies.

In any case, no  pinning due to the Co is observed until the cobalt thickness increases beyond 1nm. Above this thickness,  a continuous Co film appears to form and influence the resonant behavior of the Py.  From the inset in Fig. 5a, we note that the fundamental and SSWM's resonant fields, as functions of $M_s$, acquire a negative slope. We take this theoretical dependence on $M_s$ as  a signature of dynamic pinning.

We further note that the fields and frequencies for the 10 nm Co bilayer in Fig. 5a can be accounted for with $4\pi M_s = 15080$ G for Cobalt.  This value agrees well with a SQuID determined value of $4\pi M_s = 17800$ G for a reference 10 nm thick Co film. 

Most interesting in Fig. 5b is not only the shift in resonant fields, but also the significant change in the relative amplitudes of the resonant modes.  As stated in the previous section, with a single layer of Py, the amplitude of the first SSWM is relatively constant with respect to the fundamental mode at a constant applied field.  However, as shown in Fig. 4b, the bilayer film with just 10 nm of Co on the single Py layer has a significantly different distribution of relative amplitudes. The relative amplitude $r_i$ for a i-th mode is calculated as a ratio of its amplitude to the amplitude of the fundamental mode. This removes the effect of a decreasing precessional angle with a larger applied field.  The first SSWM increases in amplitude as the applied field increases, so much so in fact that it has a larger intensity than the fundamental mode. This effect is clearly seen for all samples with Co thicknesses greater than 5 nm, and for the whole range of Py thicknesses studied (40-91 nm).

The theoretical intensities in Fig.5b calculated using the theory in \cite{kostylev} are in good qualitative agreement with experiment. Our theory treats the microwave transducer field as absolutely homogeneous in the film plane. With this and other simplifications used in the theory, a better quantitative agreement is not to be expected. Furthermore, the theoretical intensities strongly depend on a number of material parameters, in particular on layer conductivities and the values the Gilbert magnetic damping parameters for the layers. This makes the task of optimal fitting somewhat complicated, as one has to fit all curves for intensities (Fig. 5b) and all for resonant fields (Fig. 5a) with the same set of parameters simultaneously. No attempt was made to obtain the optimal fit, as the most important task in the calculation in Fig. 5b was to show that the theoretical curves  exhibit the same behavior as in the experiment. The calculated relative intensities for SSWMs increase with frequency and, like the experimental data, reach a maximum at higher frequencies (not shown in the graph, as the theoretical maximum for the 1st SSWM is at about 18 GHz). Moreover, the theory fully explains the difference in responses for Si/Py/Co and Si/Co/Py systems (see Fig. 3 in Ref. \cite{kostylev}). Thus we conclude that the eddy currents induced in the bi-layer films by incident microwave fields give a major contribution to the broadband FMR response.

Additional measurements were carried out using the wide microstrip transducer and a hollow waveguide. As follows from Ref. \cite{silvester}, the microwave field of a 1.5 mm-wide microstrip transducer should be homogeneous above the transducer at distances less than 1.5 mm from the surface. This allows taking measurements with a sample placed on the transducer with its Si-substrate (0.5 mm thick) facing the transducer. A representative result of such a measurement with a Si/Py/Co sample is shown in Fig. 7. With the film facing the transducer, the response is Type A. The response changes to Type B when the sample is placed with the substrate facing the transducer. This is consistent with the above measurements on Si/Co/Py structures and the theoretical predictions given in Ref. \cite{kostylev}.

Following Appendix B, Case A in Ref. \cite{kostylev} the same effect of swapping response types should be seen for \textit{propagating} plane electromagnetic waves incident normally onto the film surface. We tested this using a hollow metallic waveguide of rectangular cross-section to form conditions for the normal incidence. The films are placed in the cross sectional plane of the waveguide. The measurements are made in reflection, so that the parameter S11 \cite{Counil} is obtained. The samples fill about one half of the the waveguide cross-section. For this reason one has to expect a microwave field incident on the rear sample surface that includes contributions from diffraction around edges of the sample. The presence of this diffracted field is confirmed by VNA measurements. When a sample is inserted into the waveguide, the transmission characteristic $S21(f)$ acquires a non-monotonic dependence on frequency $f$ because of partial standing wave resonances formed in the waveguide.  $S21$ varies from -6db to -12dB depending on $f$. Our field resolved measurements are carried out in a local minimum of $S21(f)$ in order to reduce effects of the diffracted microwave field at the rear sample surface.

Representative data are displayed in Fig. 8. We find the same tendency as in the microstrip experiments: responses of Type A are obtained when the Co layer of any structure faces the incident flux, and responses of Type B are obtained when the Py layer faces the incident flux.

\vspace{1cm}
\textbf{6. Discussion}

Here we give basic elements of the theory in a broader context of previous studies for eddy-current effects for conducting films. Details of the theory can be found in Ref. \cite{kostylev}. The essential result of the theory is that conditions for a highly inhomogeneous microwave magnetic field are formed in the microstrip broadband FMR geometry due to microwave screening by eddy currents in conducting samples. Microwave screening in conducting films thinner than the microwave magnetic skin depth \cite{Almeida}, hereby referred to as sub-skin depth films "SSDF", may strongly affect broadband FMR measurement results. In summary, manifestations of microwave screening for SSDF are as follows:

(i) the response of conductive multilayers may strongly depend on layer ordering with respect to the microwave transducer location; (ii)  extremely large amplitudes of high order standing spin wave modes can be observed in some multilayers; and (iii) the response of these systems can be strongly frequency dependent.

Experimental data presented here are in the full agreement with these predictions.

 We note that to some extent the driving of SSWM discussed here is similar to efficient excitation of high order FMR modes by the microwave electric field observed by P. Wolf \cite{Wolf}. In both cases, the excitation of SSWM depends upon inhomogeneous fields. In Wolf's experiment, a conducting film was placed on a hole in a cavity wall. Efficient excitation of the inhomogeneous SSWM resonances was observed and can be understood as follows. The microwave electric field across the hole drives a current in the sample. The current creates a microwave Oersted field which is anti-symmetric across the sample thickness. This field can be approximated by  ${h}_x(y) \propto y/L-0.5$. This highly inhomogeneous magnetic field efficiently excites the SSWM resonances. Importantly, this experiment clearly demonstrates that magnetisation precession is driven by the total microwave magnetic field to which microwave currents in the sample contribute.

In Ref.\cite{kostylev} it was noted that an external in-plane microwave magnetic field $\bf{h}_e$ applied to a film medium is necessarily accompanied by an in-plane curling electric field. This electric field should induce a microwave current in the sample whose Oersted field  $\bf{h}_{Oe}$ adds to the external microwave magnetic field. For thick samples one recovers the magnetic skin depth effect and the amplitude of the total field $\bf{h}_t=\bf{h}_e+\bf{h}_{Oe}$ falls off exponentially with the distance from the sample surface facing the incident field flux. Such a thick film geometry has been discussed in the past \cite{Almeida, Ament}.

It turns out that effects are striking in the case of thin films also. Indeed, for a thin film with thickness less than the SSDF, the total microwave magnetic field decays more strongly than exponentially. For Py of a thickness larger than 30 nm,  in contact with an adjoining media with a high characteristic impedance $z_0\geq 50$Ohm, the field is negligible at the rear film surface (see Eq.(44) and Fig. 7 in Ref.\cite{kostylev}).
The derivation of this result is not trivial, and details can be found in Ref. \cite{kostylev}. (See also Eqs.(3-4) in the present paper.)

It is important to note that the microwave shielding effect is not seen in the cavity FMR measurements, as the microwave magnetic field of the cavity is incident on both SSDF surfaces (see Eq.(4.1) in Ref.\cite{wolfram}). As a result the total microwave magnetic field inside an SSDF sample is close to homogeneous. If there is no magnetization pinning at the film surfaces,  the fundamental mode only displays an FMR response. Intensities for higher-order SSWMs are observable in the cavity FMR only for samples with a very high level of asymmetry; e.g. with spins almost pinned at one of the film surfaces.

The enhanced inhomogeneity originates from an extension of the classical skin depth effect to samples of finite thicknesses (see \cite{Bauer} and references therein). The phase of the back-reflection of the total microwave magnetic field from the boundary between two media with a large difference in values of electric conductivity is important for such samples. The total microwave magnetic field incident on the rear surface interferes destructively with the back reflected field. As a result, the usual skin-depth law $\left|\textbf{h}_t(y)\right| \propto exp(-\sqrt{(i\sigma \omega)}y)$ which is valid for for the half-space is modified. For a single layer of thickness $L$ the expression is ${h}_{tx}(y) \propto sinh(\sqrt{(i\sigma \omega)}(y-L))$, where $y=0$  is the coordinate for the film surface facing the transducer and and $y=L$ is for the rear film surface. For films which are much thinner than the classical microwave skin depth this expression reduces to a linear function ${h}_{tx}(y) \propto (y-L)/L$.  From this formula one sees that the total microwave magnetic field inside the samples is indeed highly inhomogeneous and strongly asymmetric.
Since  magnetization precession is driven by the total field, conditions are thereby formed for efficient excitation of non-uniform eigenmodes of precession. If eigenmodes of the system lack inversion symmetry the SSDF broadband FMR response will depend on layer ordering with respect to the direction of the incident microwave flux.

For single films absorption by the first SSWM with odd symmetry is most strongly affected. For insulating samples with unpinned surface spins this mode is not seen in FMR spectra. Our experiments demonstrate that the situation is quite different for conducting films, and this mode can even provide the largest response. We note also that efficient excitation of high order modes was previously seen in experiments on a thicker sample in Ref.\cite{fetisov} (see Fig.5 in that paper).

Even more pronounced effects can be expected for bilayers. Calculated mode profiles for the dynamic magnetization and the total field are shown in Fig. 9 for the 10 nm Co bilayer. The left panels of this figure are for the Si/Py/Co structures with the Co layer facing the transducer. The right panels are for Si/Co/Py with the Py layer facing the transducer.

The fundamental mode displays a highly inhomogeneous dynamic magnetization across the Py, and has a minimum at the interface (see Fig. 9a and c). This minimum is due to partial dynamic pinning of magnetization at the Py/Co interface. The first high order SSWM is also strongly affected by pinning, and can be decomposed into a combination of the Co fundamental mode and the 1st Py SSWM. The broadband FMR response of a resonant mode can be approximated by an overlap integral composed of the SSWM amplitude and ${h}_{tx}(y)$.

The total microwave magnetic field in a bilayer is also described by a linear function with a discontinuity at the magnetic interface where the slope scales with layer conductivity. The effect is seen in the profiles shown in the panels b and c of Fig. 9.  Comparing the left and the right panels, one sees that the overlap integral of the fundamental mode profile with the total field is clearly dependent on the layer ordering. Note however that the overlap integral for the first SSWM is only weakly dependent on the layer ordering. We conclude from this that the coupling of the fundamental mode to the total field is efficient for Si/Co/Py structures, and it should be the dominant feature in an absorption spectrum. The response of the fundamental mode in the Si/Py/Co bilayers is weakened and becomes comparable with the response of the 1st exchange mode (whose response is much less dependent on the layer ordering).

It is also worth noticing that in Figs. 3 and 5 we do not convert the values of the measured sample response $S21(H)/S21_0(H)$ into the scalar magnetic permeability as is often done \cite{Devolder2}. The reason for this is as follows. First, the permeability one obtains in this way is an effective scalar permeability but not the tensor of the microwave magnetic permeability \cite{Gurevich}. Secondly, the effective permeability value which is extracted from the experimental data is found to a constant \cite{Devolder1, Crawford}. This constant depends on the geometry of the stripline waveguide and should be obtained from theory which should be constructed separtely for each type of transducer. On the contrary, the values of the measured $S21(H)/S21_0(H)$ allow extraction of a physically meaningful parameter of the transducer complex radiation resistance $Z_r=2z_0 l \ln((S21/S21_0))$, where $z_0$ is the characteristic impedance of the transducer and $l$ is the sample length along the transducer. The imaginary part of the effective permeability scales linearly with $Re(Z_r)$ and its real part scales with $Im(Z_r)$. We found that $\mid Im(Z_r) \mid << Re(Z_r)$ for all our experimental data. This is in good agreement with the theoretical result shown in Fig.1 of Ref.\cite{kostylev}.  Thus $S21(H)/S21_0(H)\approx Re(S21(H)/S21_0(H))$ and $Re(Z_r) \approx 2z_0 l \ln(Re(S21/S21_0))$.

Note that $Z_r$ values extracted from the experiment are absolute, and therefore can be used for extraction of film material parameters using the existing theory in Ref.\cite{kostylev} or similar. Furthermore, the $Re(S21(H)/S21_0(H))$ values off-resonance represent contribution to $Z_r$ from eddy-current losses. For metallic samples this nonmagnetic contribution is much larger than the precessional magnetic one as seen from Figs. 3 and 5. For different samples the off-resonance transmission $Re(S21(H)/S21_0(H))$ varies from 0.2 to 0.5 (i.e. from -14 to -6 dB), but the resonance contribution is less than 1 percent  (0.1 dB) of the off-resonance value in all panels.

Finally, this effect should be observed for a number of different excitation geometries provided that the microwave field flux is incident on the bi-layer structure from one surface only. The present experiment with the hollow waveguide (Fig. 9) is in full agreement with this prediction.

\vspace{1cm}
\textbf{7. Conclusion: comparison of different measurement techniques}

Unlike what is observed for single films, we find a strong frequency dependence of the amplitude of the first standing spin wave mode relative to the fundamental mode amplitude in stripline response for bilayers.  Most significantly, the response amplitude is strongly dependent on the ordering of layers with respect to a stripline transducer.  As discussed previously, quantitative analysis of the observed mode amplitudes can be made in terms of screening by eddy currents existing in the metallic films with thicknesses below the microwave magnetic skin depth. This effect, as illustrated by our examples of exchange coupled Co/Py bilayers, is useful for studying buried magnetic interfaces and exchange effects in conducting structures.

We conclude by discussing the reasons why efficient excitation of high order standing spin wave modes was observed with  coplanar and stripline transducers, but not with cavity FMR or inelastic light scattering. We begin the discussion by noting that according to Eq. 2, it would seem that conductivity will affect the microwave response for $\mu_0\sigma \omega$ values comparable with $k^2$, i.e. for spin waves with wavelengths $2\pi/k$ comparable to the microwave skin depth in the material $\sqrt{2/(\mu_0\sigma \omega)}$. However, as shown by Almeida and Mills \cite{Almeida}, the range of affected in-plane wavenumbers is considerably larger. The important parameter turns out to be the microwave \textit{magnetic} skin depth $l_{sm}$ which can be considerably smaller than the classical skin depth $l_{sc}=\sqrt{2/(\mu_0\sigma \omega)}$, especially at frequencies and applied fields close to those for the in-plane ferromagnetic resonance.

The dynamic fields are related by the microwave magnetic susceptibility tensor $\hat \chi$, defined by  $\textbf{m}_k=\hat \chi \textbf{h}_k$. The susceptibility can be found from the linearised Landau-Lifschitz-Gilbert equation \cite{Gurevich}. This, together with Eqs. (1) and (2) form a system of equations with solution
$\textbf{m}_k,\textbf{h}_k \propto exp( \pm Q y)$. The out-of-plane wavenumber $Q$ is larger than the corresponding insulating value ($k$), and is also larger than the quantity $\sqrt{k^2i \mu_0 +\sigma \omega}$ appearing in Eq.(2). The imaginary component of $Q^2$, and the actual magnetic skin depth, are  "amplified"' by the off diagonal susceptibility (or magnetic gyrotropy) $\chi_a$:
$l_{sm}=l_{sc}/\sqrt{\mu_V}$, and $Q=\sqrt{k^2 \mu_0-i\mu_V\sigma \omega}$, where $\hat \mu = \hat 1+ 4\pi \hat \chi$ and $\mu_V= \frac{\mu^2+\chi_a^2}{\mu}$ \cite{Almeida}.

Ferromagnetic resonance in our geometry represents homogeneous precession $k=0$. In the absence of magnetic losses at the resonance frequency, the diagonal component of the permeability tensor vanishes ($\mu=0$). However the off diagonal component $\mu_a$ responsible for gyrotropy does not vanish at resonance. This results in divergence of $\mu_V$ at resonance.

In real materials $\mu_V$ is bounded due to  magnetic losses, but nevertheless $l_{sm}$ is nearly an order of magnitude greater than $l_{sc}$ for  Permalloy films. For $k>0$ the conductivity contribution to $Q$ becomes less important, but still large. This effect is shown in Fig. 2a. From this figure one sees that spin waves with in-plane wave numbers up to $40000$ rad/cm are affected by the conductivity.

The consequences for resonant absorption due to this enhanced skin depth appear when one considers the stripline geometry. Spin waves are excited within an area in which the transducer's magnetic field is largest. They travel out of this region, mostly in directions perpendicular to the transducer axis. It is known that spin waves are excited by stripline transducers resonantly, so that a microwave field of the frequency $\omega$ excites a spin wave with the same frequency (see e.g.\cite{ganguly, dmitriev-kalinikos, dmitriev, tschneider}) with in-plane wavenumber $k$ determined by the spin wave dispersion  $\omega(k)$. The amplitude of an excited spin wave with wavector $k$ is proportional to the amplitude of the corresponding spatial Fourier-component of the transducer's microwave magnetic field $\textbf{h}_{ek}$.

The spectrum of Fourier components for the microwave field from the microstrip antenna (Fig.1b) is $\left| \textbf{h}_{ek}\right| \propto sin(kw/2)/(kw/2)$ (see Eq.(38) in \cite{kalinikos81}). The first zero of this function is located at $k=2\pi/w$. The Fourier spectrum of the microwave field of coplanar transducers (Fig.1a) has a more complicated shape with several minor lobes grouped together (see e.g. Fig.{3} in \cite{Kim}). It is usually assumed that the wavenumber bandwidth excited by a coplanar transducer is given by the width of the first major lobe of the Fourier spectrum (as shown in Fig. 1 in \cite{Kim} this is between $k=0$ and $200$ rad/cm). From this we conclude that for both types of transducers the wavenumber range for spin wave excitation is from  $k_{min}=0$ to $k_{max}=2\pi/w_{char}$, where $w_{char}$ is the characteristic width of the transducer. For the microstrip transducers, $w_{char}$ coincides with the width of the microstrip $w$ in the $x$ direction. For coplanar transducers $w_{char}=w+2\Delta$, where $w$ is the width of the center conductor and $\Delta$ is the separation of the central conductor from a ground half-plane \cite{dmitriev, Kim}.

Transducers having a characteristic width of $2-5$ micron were used in experiments on traveling spin waves reported in \cite{bailleul, bailleul-science, bailleul1, demidov, demidov1, fetisov}. These correspond to wavenumbers $k_{max}=10^4-3\cdot10^4$ rad/cm. Typically, a single-layer $30-40$ nm-thick Permalloy film was utilised (except for \cite{fetisov}, where the film was $200$ nm thick). The calculated results shown in Fig.2  are for a $40$ nm thick film for comparison. One sees that that conductivity effects in $Q(k)$ (Fig.2a) appear throughout the entire transducer wave number range. Furthermore, in Brillouin light scattering experiments the accessible wavenumber range extends to $2.5\cdot10^5$ rad/cm (see e.g. \cite{Hillebrands}), hence a large part of the accessbile wavenumber range is affected also. 

Nevertheless, zero conductivity models the frequencies work exceptionally well for travelling wave experiments, such as Brillouin light scattering, and for coplanar-transducer driven travelling spin waves experiments on 30-40nm thick conducting films \cite{bailleul1}. The reasons are very illuminating, and provide insight into the fundamental nature of conductivity effects.

Observable quantities in travelling wave experiments are spin wave dispersion (usually from Brillouin light scattering) and spin wave amplitudes (most easily measured using transducers techniques). As pointed out in \cite{Almeida}, conductivity affects strongly the out of plane wavenumber. However this produces a relatively weak modification of the spin wave dispersion. Indeed, one sees in Fig. 2b the dispersion calculated with nonzero conductivity agrees very well with the dispersion calculated with zero conductivity.

In Section 2 we have already discussed  the effect of the finite film thickness and found that for SSDF films the efficient shielding disappears at a $k=200$ rad/cm (Fig. 2b). In the case of transducer studies of travelling spin waves, the wave number range from 0 to 200 rad/cm represents a small portion of the full transmission characteristic accessible. The remainder of the range is expected to be that of a magnetic insulator. The smallness of the affected wavenumber range is the main reason why any evidence of the presence of the efficient microwave shielding is not significant in the travelling wave experiments with SSDF films.

We can now understand the essential differences for typical broadband FMR experiments \cite{Silva, Crawford}. The free spin wave propagation path depends on magnetic losses in the material and on the sample thickness. Film thickness are typically below 100 $\mu$m, and the transducers are orders of magnitude larger (in our case the width of the microstrip is $w$=1.5 mm, and the width of the coplanar waveguide at $w$=0.35 mm, with $\Delta$=0.6 mm). These systems therefore fall in the limiting broad case case $k_{max} << k_s$.Typically the transducer width is chosen such that it is smaller than the free spin-wave propagation path $l_{SW}=\alpha \omega /V_g$ in order to ensure a quasi-homogeneous microwave field in the film plane (here $\alpha$ is the Gilbert damping constant and $V_g$ is the spin wave group velocity). In this way the travelling spin wave contribution \cite{Counil} to the absorption linewidth is minimized. However, for this geometry $k_{max}$ is virtually zero, so that eddy currents should have a major impact on the magnetization precession. Note that shielding cannot be observed directly, however as shown in \cite{kostylev}, looking at exchange effects with broadband FMR allows one to determine clearly the presence of efficient eddy current screening.

In conclusion, microwave magnetic dynamics in conducting films driven by wide stripline transducers is quite different from that observed in the traveling wave experiments and cavity FMR experiments on the sub magnetic skin depth films. Neither travelling wave experiments nor  cavity FMR results are noticeably impacted by the eddy currents in the sample. For the traveling-wave experiments $\textbf{h}_{Oek}$ and $\textbf{h}_{prk}$ are negligible compared with $\textbf{h}_{dk}$. For the cavity FMR, although $\textbf{h}_{Oek}$ and $\textbf{h}_{prk}$ are present and large,  they are quasi-homogeneous across $L$ and thus simply renormalise the homogeneous driving field. Furthermore, experiments on Brillouin light scattering on thermal magnons (see e.g. \cite{Hillebrands}) do not reveal any noticeable impact of the eddy currents either, as the only possible eddy current contribution to the total field of thermal magnons is $\textbf{h}_{prk}$ which has a negligible effect on magnon dispersion.

\vspace{1cm}
\textbf{Acknowledgments}
Australian Research Council support through Discovery Projects and Postgraduate Awards is acknowledged. We also acknowledge support from the University of Leeds, the EPSRC, and from the University of Western Australia.

\newpage

FIGURE CAPTIONS

Fig. 1. Cross-sections of the coplanar (a) and the microstrip (b) broadband FMR transducers with a sample on top.

Fig. 2. Solution of the exchange-free equations. (a): Red solid line: real part of the out-of-plane wavenumber $Q$. Blue dashed line: its imaginary part. Green dotted line: assymptotics for $Q$ for large wavenumbers ($Q=k$).
(b): Thin red solid line: Amplitude of the microwave field $h_{xk}$ at the rear film surface $y=L$. Thin blue dashed line: amplitude of $h_{xk}$ at the front film surface $y=0$ on which the microwave flux is incident. Green dotted line: assymptotics for both for large wavenumbers $k$. Thick black solid line in both panels: Dispersion of the Damon-Eshbach wave in an insulating film, coincides with graphical accuracy with the result obtained including the electric conductivity.
Parameters of calculation: film thickness: 40 nm, film saturation magnetisation $4\pi M_s$=10000 G, applied field is 2000 Oe, film conductivity is $4.6 \cdot 10^6$ Sm/m.

Fig. 3. Microwave broadband FMR absorption data for single-layer Permalloy films. Microstrip transducer is 1.5 mm in width. Left-hand panels (a-d): driving frequency is 7.5 GHz. Right-hand panels (e-h): 18 GHz.
Film thicknesses: (a) and (e): 30 nm; (b) and (f): 40 nm; (c) and (g): 60 nm; (d) and (h): 80 nm.

Fig. 4. Cavity FMR (Panels (a-c), dashed line) and Brillouin light scattering (BLS) data (Panel (d)) in comparison with broadband FMR data (Panels (a-c), solid lines) for the single layer films with different thicknesses. (a): 40 nm; (b): 60 nm; and (c): 80nm. Driving frequency is 9.47 GHz. Broadband transducer: the same 1.5mm-wide microstrip. Cavity: Varian-4 ESR spectrometer cavity. BLS data are taken for the 60nm-thick film at an incidence angle of 5 angle degree and in an applied field of 500 Oe. "F" in Panel (d) indicates the fundamental ("Damon-Eschbach") BLS peak, "SSWM1" is the first (odd-symmetry) standing spin wave mode, and "SSWM2" is the second (even-symmetry) standing spin wave mode.

Fig. 5. (a) Resonant mode frequencies as a function of applied bias field. Inset: calculated resonant fields as a function of saturation magnetization $4\pi M_s$ (given in Gauss) for cobalt. Driving frequency for the inset is 13.47 GHz. Solid lines in the inset: calculation; dashed lines: measured values for the bi-layer sample.  (b): Relative mode intensities.

Coplanar transducer with a central conductor 0.3 mm in width is used. Black triangles: single layer 74nm-thick film.
Red dots: bi-layer film (Py:74nm, Co:10nm). Black dashed lines: fits for the single-layer film. Blue solid lines:  fits for the bi-layer film. f: fundamental mode. 1-3:  Standing spin wave modes with respective numbers.

Fig. 6. Broadband FMR absorption for bi-layers. Microstrip transducer is 1.5 mm in width. Left-hand panels (a-d): driving frequency is 7.5 GHz. Right-hand panels (e-h): 18 GHz.

(a) and (e): Single-layer Permalloy films (Si/Ta/Py/Ta, given here for comparison); solid line: 80 nm thick; dashed line: 40 nm thick;

(b-d) and (f-h) are bi-layer responses. Red solid lines are for Si/Ta/Py/Co/Ta and dashed blue lines are for Si/Ta/Co/Py/Ta structures. In all figures Cobalt layer is 10nm thick.

(b) and (f): Permalloy thickness is 40 nm;
(c) and (g): 60 nm;
(d) and (h): 80 nm.
Inset to panel (d): cavity FMR data for the respective Si/Py/Co structure (80nm of Py, 10 nm of Co.) The cavity data were taken at 9.47 GHz.

Fig. 7. Microwave broadband FMR absorption for the bi-layer from Fig. 5. The 1.5mm-wide microstrip transducer is used. Solid line: film faces the microstrip. Dashed line: film substrate faces the microstrip.

Fig. 8. Hollow waveguide data. Frequency is 9.5 GHz. (a): response of Si/Ta/Py[80nm]/Co[10 nm]/Ta structure. (b): response of Si/Ta/Co[10 nm]/Py[80 nm]/Ta structure. Red solid line: Film facing the incident flux. Blue dashed line: substrate facing the incident flux. All measurements were done in reflection (S11 parameter was measured).

Fig. 9. (Color online) Calculated profiles of the dynamic magnetization and of the total microwave magnetic field for the bi-layer Py[74nm]/Co[10nm]. (a) and (c) in-plane component of dynamic magnetization $m_x$ at 7.5 GHz.  (b) and (d): total microwave magnetic field.  Left panels: Permalloy layer faces the transducer. Right panels: cobalt layer faces the transducer. Blue lines: phase for the first standing-spin wave mode of the stack (right axes).

\newpage

\begin{figure}
	\centering
		\includegraphics{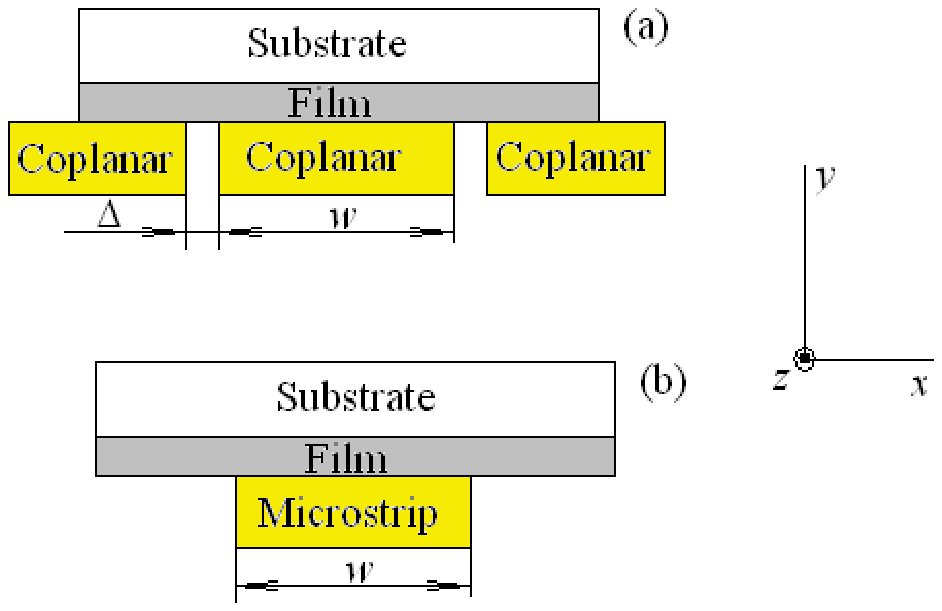}
\end{figure}

\newpage

\begin{figure}
	\centering
		\includegraphics{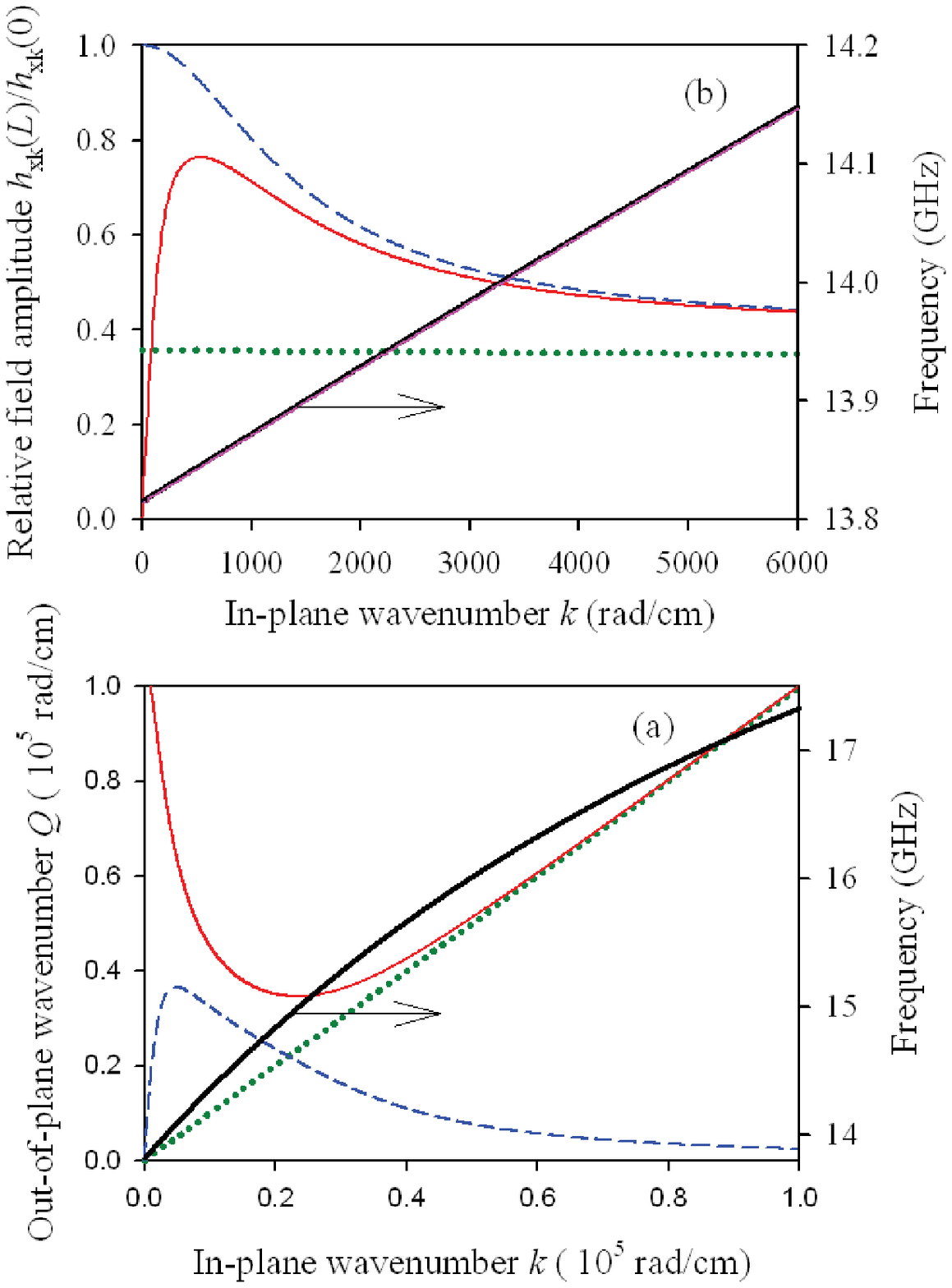}
\end{figure}

\newpage

\begin{figure}
	\centering
		\includegraphics{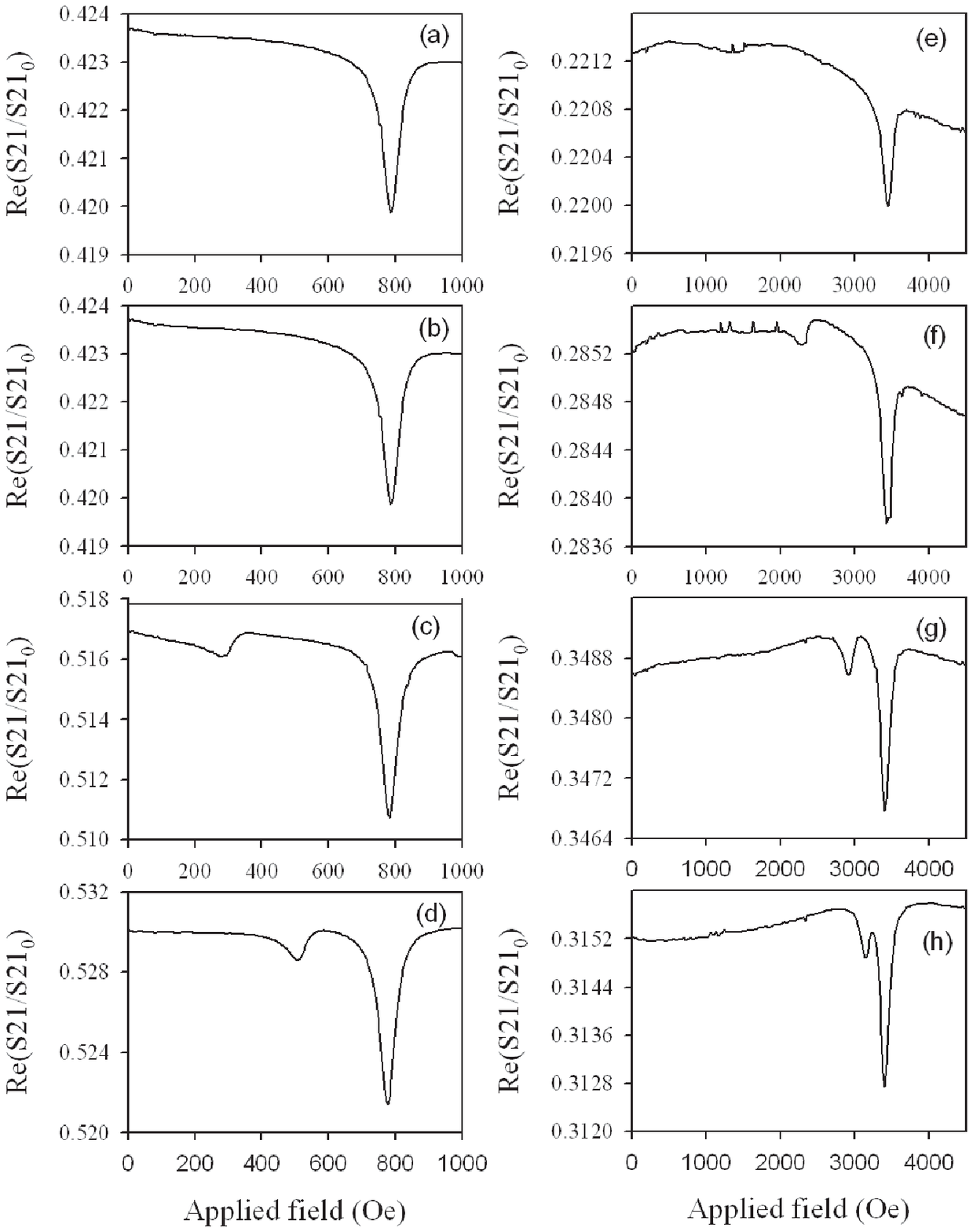}
\end{figure}

\newpage

\begin{figure}
	\centering
		\includegraphics{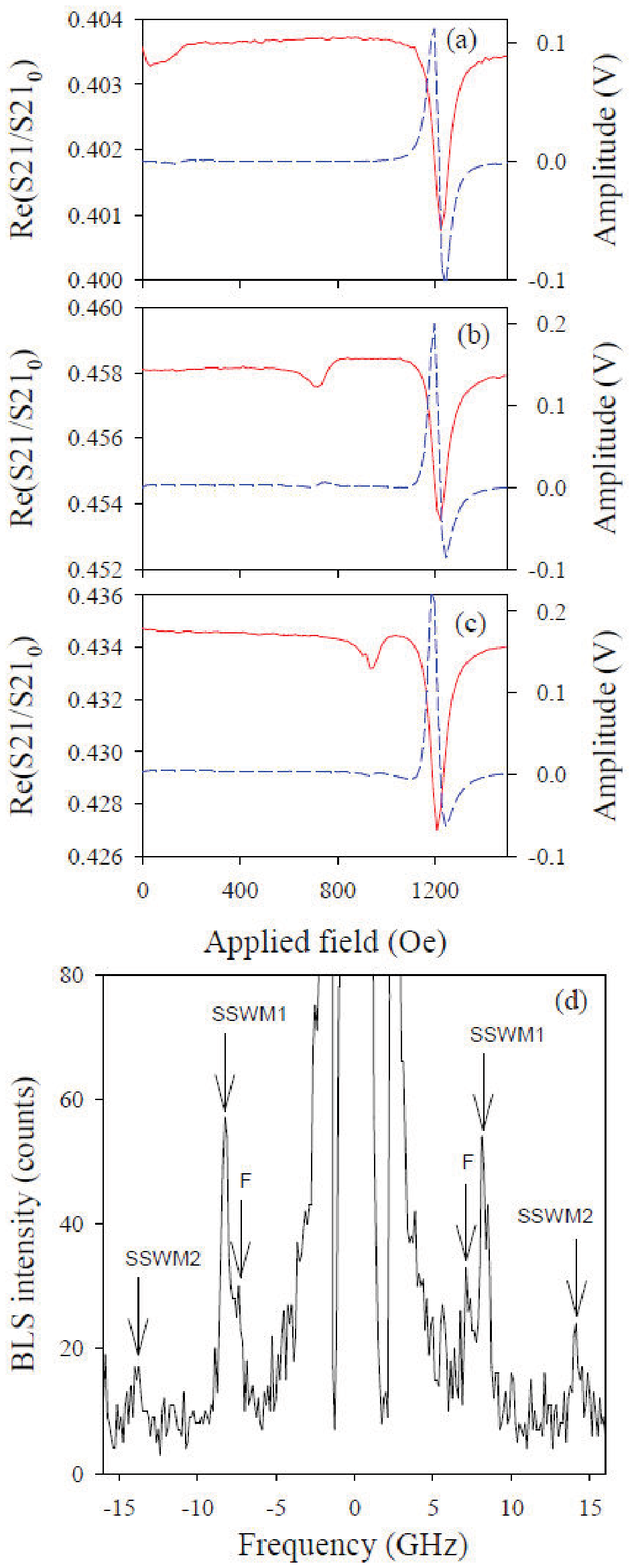}
\end{figure}

\newpage

\begin{figure}
	\centering
		\includegraphics{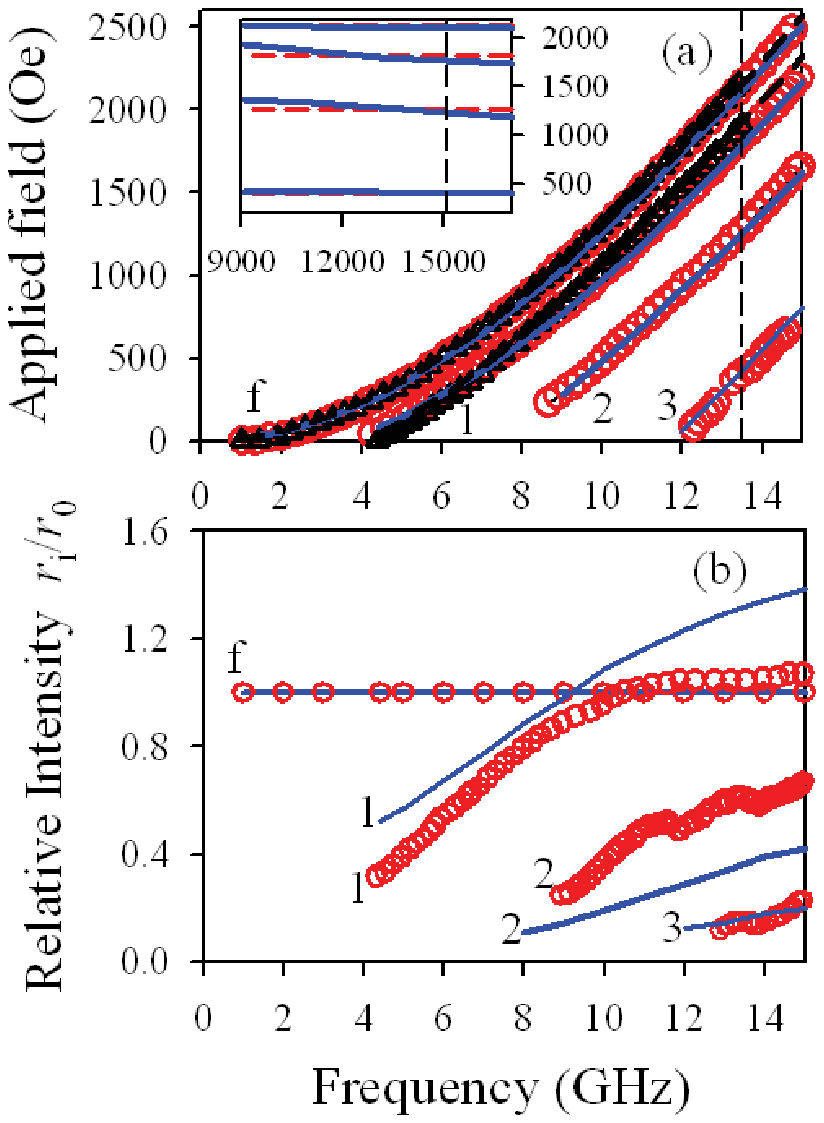}
\end{figure}

\newpage

\begin{figure}
	\centering
		\includegraphics{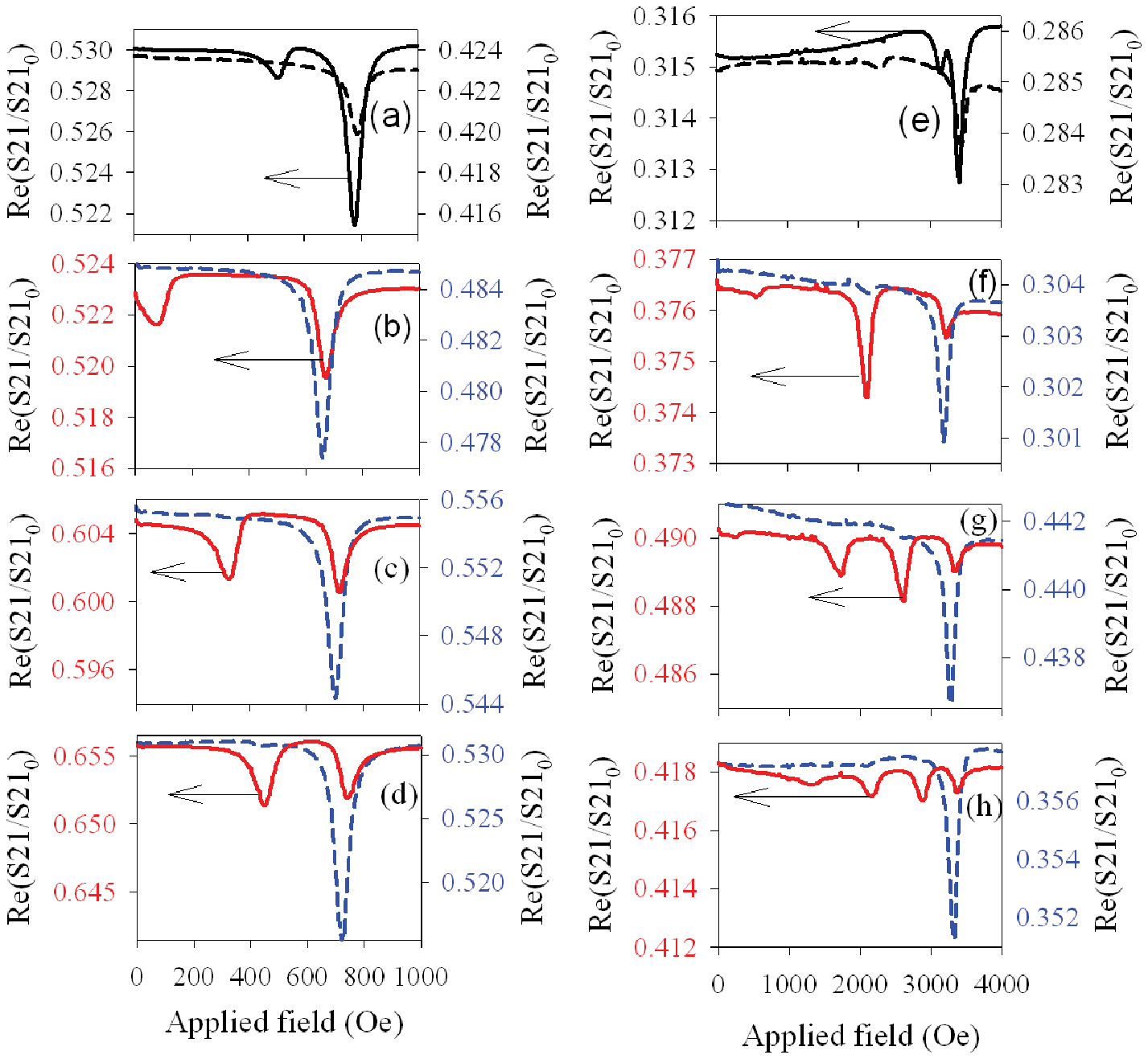}
\end{figure}

\newpage

\begin{figure}
	\centering
		\includegraphics{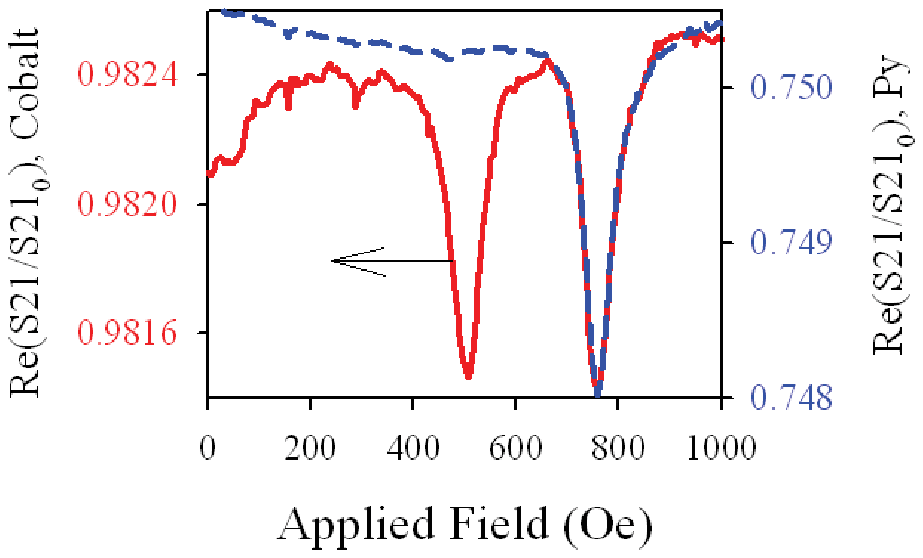}
\end{figure}

\newpage

\begin{figure}
	\centering
		\includegraphics{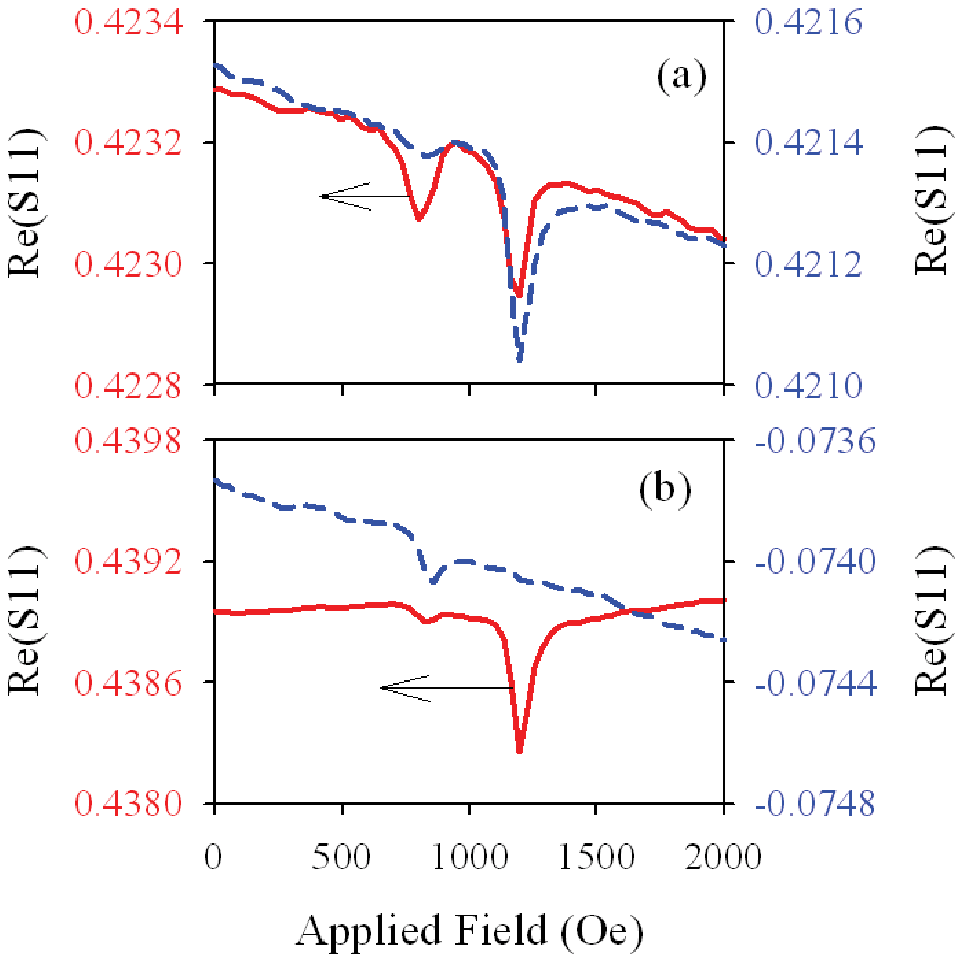}
\end{figure}

\newpage

\begin{figure}
	\centering
		\includegraphics{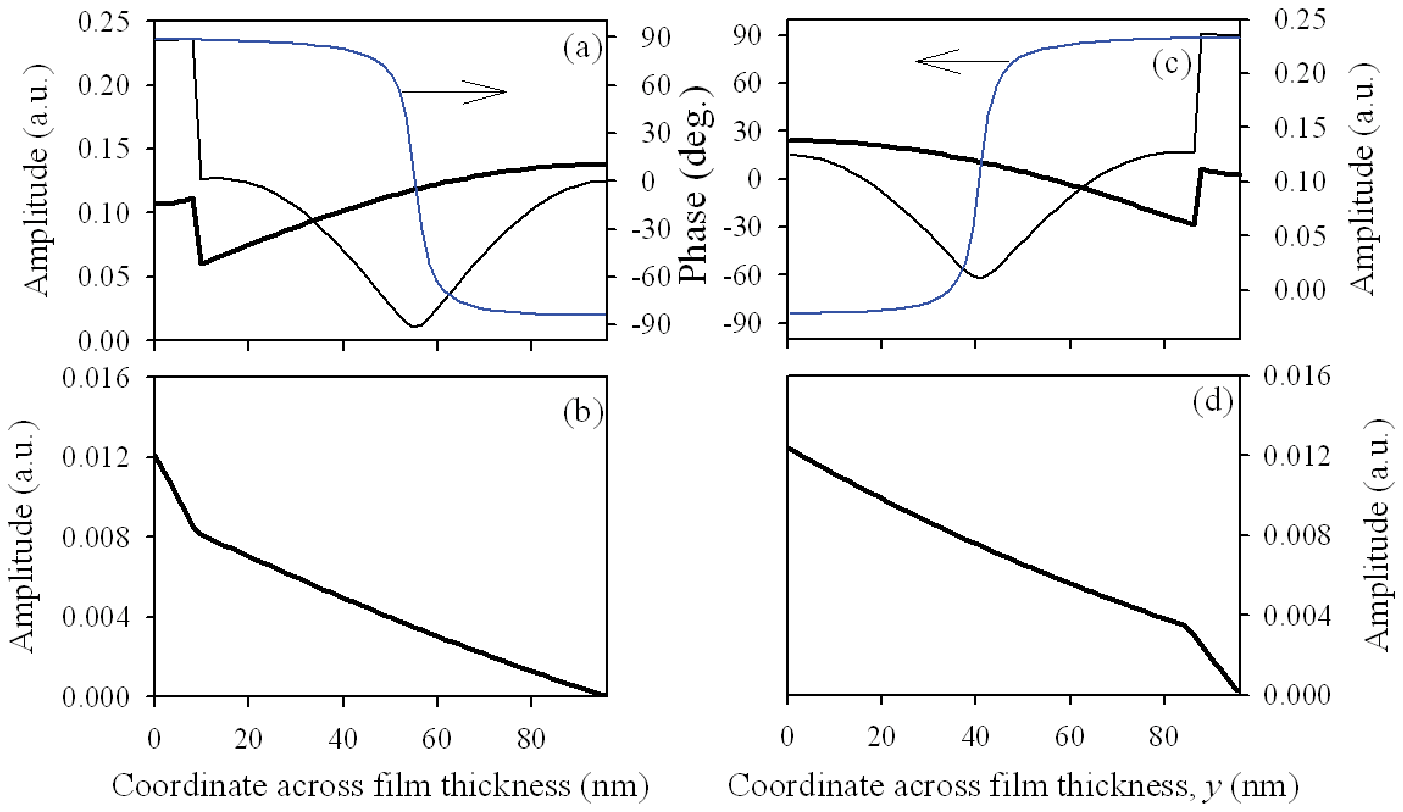}
\end{figure}

\end{document}